\documentclass{jfm}

\usepackage[normalem]{ulem}
\usepackage{hyperref}
\hypersetup{
    colorlinks = true,
    urlcolor   = blue,
    citecolor  = black,
}
\newcommand{\RomanNumeralCaps}[1]
\linenumbers
\usepackage{amsmath}
\usepackage{amssymb}
\usepackage{amsfonts}
\usepackage{fdsymbol}
\usepackage{gensymb}

\usepackage{textcomp}
\usepackage{wasysym}
\usepackage{siunitx}

\usepackage{graphicx}
\usepackage{epstopdf, epsfig}
\usepackage{float} %use [t] or [t] to force inline image

\restylefloat{table}
\restylefloat{figure}
\usepackage{subcaption}

\usepackage[dvipsnames]{xcolor}
\definecolor{mo}{RGB}{108, 172, 228}
\definecolor{aa}{RGB}{170, 21, 27}
\definecolor{nm}{RGB}{0, 38, 84}
\title{An experimental study on the settling velocity of inertial particles in different homogeneous isotropic turbulent flows}

\author{Am\'elie Ferran\aff{1, 2} \corresp{\email{amelie.ferran1@univ-grenoble-alpes.fr}}, Nathanaël Machicoane\aff{1}, Alberto Aliseda\aff{2}, Martín Obligado\aff{1}
}
\affiliation{
\aff{1} Universit\'e Grenoble Alpes, CNRS, Grenoble-INP, LEGI, F-38000 Grenoble, France 
\aff{2} Department of Mechanical Engineering, University of Washington, Seattle, Washington 98195-2600, USA}

\begin{document}
\maketitle 
\begin{abstract}
We propose an experimental study on the gravitational settling velocity of dense, sub-Kolmogorov inertial particles under different background turbulent flows. We report Phase Doppler Particle Analyzer measurements in a low-speed wind tunnel uniformly seeded with micrometer scale water droplets. Turbulence is generated with three different grids (two consisting on different active-grid protocols while the third is a regular static grid), allowing us to cover a very wide range of turbulence conditions in terms of Taylor-scale based Reynolds numbers ($Re_\lambda \in [30-520]$), Rouse numbers ($Ro \in   [0-5]$) and volume fractions ($\phi_v \in[0.5\times10^{-5} - 2.0\times10^{-5}]$). 

We find, in agreement with previous works, that enhancement of the settling velocity occurs at low Rouse number, while hindering of the settling occurs at higher Rouse number for decreasing turbulence energy levels. The wide range of flow parameters explored allowed us to observe that enhancement decreases significantly with the Taylor Reynolds number and is significantly affected by the volume fraction $\phi_v$. We also studied the effect of large-scale forcing on settling velocity modification. The possibility of change the inflow conditions by using different grids allowed us to test cases with fixed $Re_\lambda$ and turbulent intensity but different integral length scale. Finally, we assess the existence of secondary flows in the wind tunnel and their role on particle settling. This is achieved by characterising the settling velocity at two different positions, the centreline and close to the wall, with the same streamwise coordinate.
\end{abstract}

\begin{keywords}
%Authors should not enter keywords on the manuscript, as these must be chosen by the author during the online submission process and will then be added during the typesetting process (see \href{https://www.cambridge.org/core/journals/journal-of-fluid-mechanics/information/list-of-keywords}{Keyword PDF} for the full list).  Other classifications will be added at the same time.
\end{keywords}

%{\bf MSC Codes }  {\it(Optional)} Please enter your MSC Codes here

\section{Introduction}
Turbulent flows laden with particles are present in both environmental phenomena and industrial applications. For instance, water droplets, snowflakes and pollutants in atmospheric turbulence, sediments in rivers and industrial sprays all involve turbulent environments carrying inertial particles (\cite{Crowe1996, Shaw2003, Monchaux2012ijmf, Li2021}). Inertial particles do not follow the fluid velocity field as tracers, having their own dynamics that depend on both their finite size and their density ratio compared to that of the carrier phase.

Two phenomena resulting from the influence of turbulence on the motion of inertial particles have been widely studied: preferential concentration and modification of the settling velocity.
Preferential concentration refers to the fact that an initially uniform or random distribution of particles will form areas of clusters and voids (\cite{Maxey1987, Squires_Eaton1991, Aliseda2002, Obligado2014, Sumbekova2017}) due to the accumulation in certain regions of the turbulent flow where the hydrodynamic forces exerted by the flow tend to drive the particles.
Furthermore, settling velocity modification occurs when particles immersed in a turbulent flow have their settling speed $V_s$ altered compared to that in a stagnant fluid or laminar flow $V_T$ (\cite{wang_maxey_1993, Crowe1996, Aliseda2011}).
These two features of turbulent-laden flow are known to be linked together as the settling velocity of a particle can be increased due to an increase of the particle local concentration (\cite{Aliseda2002, Gustavsson2014, Huck2018}).

Regarding the modification of the settling velocity, multiple experimental and numerical studies have shown that turbulence can both hinder ($V_s < V_T$) or enhance the particle settling velocity ($V_s > V_T$).
While several studies have reported enhancement of the settling velocity (\cite{wang_maxey_1993, Aliseda2002, Bec2014, Rosa2016, Monchaux2017, Falkinhoff2020}), others show evidence of hindering only (\cite{Mora2021settling}) or of both types of modification (\cite{Nielsen93, Good2012, Sumbekova2016, Petersen2019}). While the nature and number of mechanisms controlling this phenomenon is still a matter of debate, several models have been proposed in the literature, sometimes even giving contradictory predictions.

Enhancement of the settling velocity can be explained by the preferential sweeping mechanism, also known as fast-tracking effect, where inertial particles tend to spend more time in downwards moving regions of the flow than in upwards flow (\cite{wang_maxey_1993}). 
Some mechanisms have been proposed as well to explain hindering. The vortex trapping effect describes how light particles can be trapped inside vortices (\cite{Nielsen93, Aliseda06}). The loitering mechanism assumes that falling particles spend more time in upward regions of the flow than downward regions (\cite{Chen2020}), while a nonlinear drag can also explain that particles are slowed down in their fall by turbulence (\cite{Good2014}). 
Models have been developed to estimate the influence of clustering and particle local concentration on the settling rate enhancement (\cite{Alipchenkov2009, Huck2018}).  

However, even in the simplified case of small, heavy particles in homogeneous isotropic turbulence (HIT) no general consensus has been found on the influence of  turbulence, through the Taylor-Reynolds number $Re_\lambda$, on the transition between hindering and enhancement. %and on the maximum of enhancement .
$Re_\lambda = u'\lambda/\nu$ is based on the Taylor microscale $\lambda$ where $u'$ and $\nu$ are the carrier phase rms (root-mean-square) of the fluctuating velocity and kinematic viscosity respectively. 
The influence of $Re_\lambda$ on the maximum of enhancement, i.e. when $V_s - V_T$ reaches its maximum, is also still under debate.
Depending on the range of $Re_\lambda$, some studies found that the maximum enhancement increases with $Re_\lambda$ (\cite{Nielsen93, Yang1998, Bec2014, Rosa2016, Wang2018}), whereas other studies show the opposite trend (\cite{Mora2021settling}).
Furthermore, a non-monotonic behaviour of $\max(V_s - V_T)$ with $Re_\lambda$ has also been reported (\cite{Yang2021}), where $\max(V_s - V_T)$ corresponds to the maximal settling velocity with respect to the terminal velocity, with both $V_s$ and $V_T$ being functions of the particle size. 

Several non-dimensional parameters have been found to play a role on the settling velocity. The dispersed phase interactions with turbulent structures are characterised by the Stokes and Rouse numbers (\cite{Maxey1987}), whereas the magnitude of turbulence excitation is quantified by the Taylor Reynolds number.
The Stokes number, describing the tuning of particle inertia to turbulent eddies turn over time, is defined as the ratio between the particle relaxation time and a characteristic timescale of the flow $St = \tau_p/\tau_k$, where $\tau_k$ has been shown to be represented by the Kolmogorov time scale $\tau_\eta$. 
The Rouse number - also known as $Sv$ - is a ratio between the particle terminal speed and the velocity scale of  turbulence fluctuations, in this case the turbulent velocity rms, $Ro = V_T/u'$. Hence, it is a competition between turbulence and gravity effects.
While all these parameters are relevant for modelling and understanding the interactions of inertial particles and turbulence, there are still no consensus even on the set of non-dimensional numbers required to do so. Furthermore, the determination of length and time flow scales relevant to the settling speed modification has also been the subject of significant discussion in the literature.
\cite{Yang1998} determined that a mixed scaling using both $\tau_\eta$ and $u'$ appears to be an appropriate combination of parameters for the present problem.
There is a general agreement that the modification of the settling velocity is a process that encompass all turbulent scales and, consistent with even single-phase HIT, a single flow scale is not sufficient to completely describe it.
It has been shown that the particle settling velocity is affected by larger flow length scales with increasing Stokes number (\cite{Tom2019}). 

Experimentally, the influence of turbulence on the particle settling velocity has been studied in an air turbulence chamber (\cite{Petersen2019} \cite{Good2014}), channel flows (\cite{Wang2018}), Taylor Couette flows (\cite{Yang2021}), water tank with vibrating-grids turbulence (\cite{Yang2003, Poelma2007, Zhou2009, Akutina2020}) and wind tunnel turbulence (\cite{Aliseda2002, Sumbekova2017, Huck2018, Mora2021settling}).
However, measuring the particle settling velocity in confined flows, such as the wind tunnel, can be challenging due to  the recirculation currents that may arise on the carrier phase.
Weak carrier phase currents in the direction of gravity can be of the order of the smallest particle velocity and impact significantly the measurements of the settling velocity, (as reported in \cite{Sumbekova2016phd, Wang2018, Akutina2020, Mora2021settling, Pujara2021, DeSouza2021}). 
\cite{Akutina2020} dealt with this bias by removing the local mean fluid velocity from the particle instantaneous velocity measurements. 

Accurate measurements of settling velocity and the local properties of the carrier-phase flow are therefore one aspect of major importance to better understand the role of turbulence on settling velocity modification. This work studies the settling velocity of sub-Kolmogorov water droplets in wind tunnel grid-generated turbulence. 
Turbulence is generated with three different grids (two consisting on different active-grid protocols while the third is a regular static grid), allowing us to cover a very wide range of turbulence conditions, with the turbulence intensity $u'/U_\infty$ ranging from 2 to 15\%, $Re_\lambda \in [34, 520]$ and integral length scales $\mathcal{L} \in [1, 15]$ cm.
Furthermore, we explore experimental realisations with similar values of $Re_\lambda$ and $u'/U_\infty$ but significantly different (by a factor 2) values of $\mathcal{L}$. This allowed us to disentangle the role of the large-scale forcing of the flow on settling velocity modifications, opening the door to expand available models to non-homogeneous flows. 

Particle settling velocity and diameter were quantified using a Phase Doppler Particle Analyzer (PDPA), as described by \cite{Mora2021settling, MorapaibaPhD} for the same facility. The present work is unique as it covers a broad range of turbulent flows, while resolving the settling velocity of particles as small as \SI{10}{\um}. Furthermore, we assess the existence of secondary flows in the wind tunnel and their role on particle settling. This is achieved by characterising the settling velocity at two different positions, the centreline and close to the wall, with the same streamwise coordinate. These measurements were complemented by hot-wire anemometry (that allows to resolve all scales flow) and a Cobra probe (a multi-hole pitot tube that resolves, in our conditions, the average and rms values of the 3D velocity vector, \cite{Obligado2022}) in single-phase conditions for all three grids studied. This approach allowed us to relate the behaviour of the smallest particles injected in the flow (\SI{10}{\um})  with the carrier phase large-scale topology.

\section{Experimental setup}
\subsection{Grid Turbulence in the Wind Tunnel}
\begin{figure}[t!]
    \centering
    \begin{subfigure}[b]{0.95\textwidth}
        \centering \includegraphics[width=\textwidth]{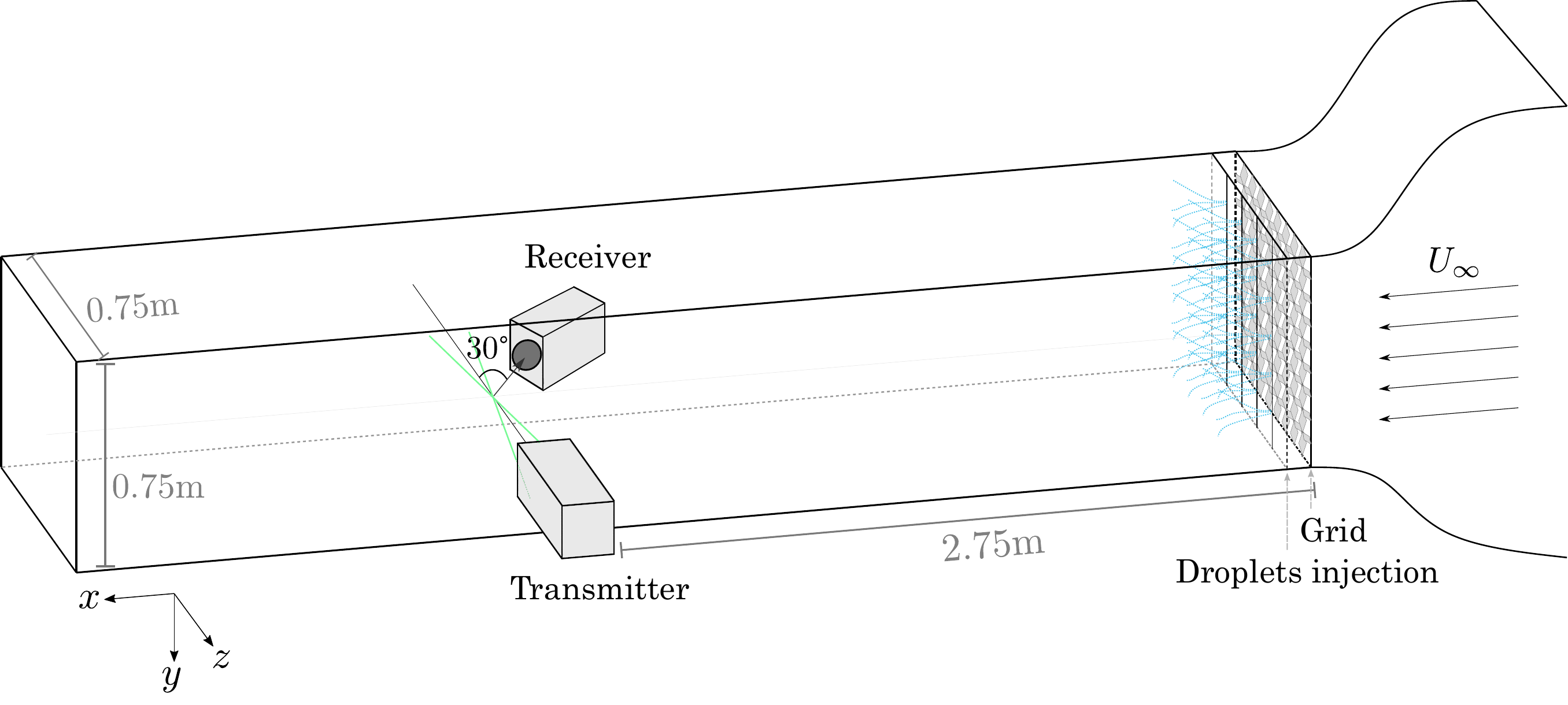}
        \caption{}
        \label{fig:WT}
    \end{subfigure}
    \\
    \begin{subfigure}[b]{0.38\textwidth}
        \centering \includegraphics[width=\textwidth, angle=-90]{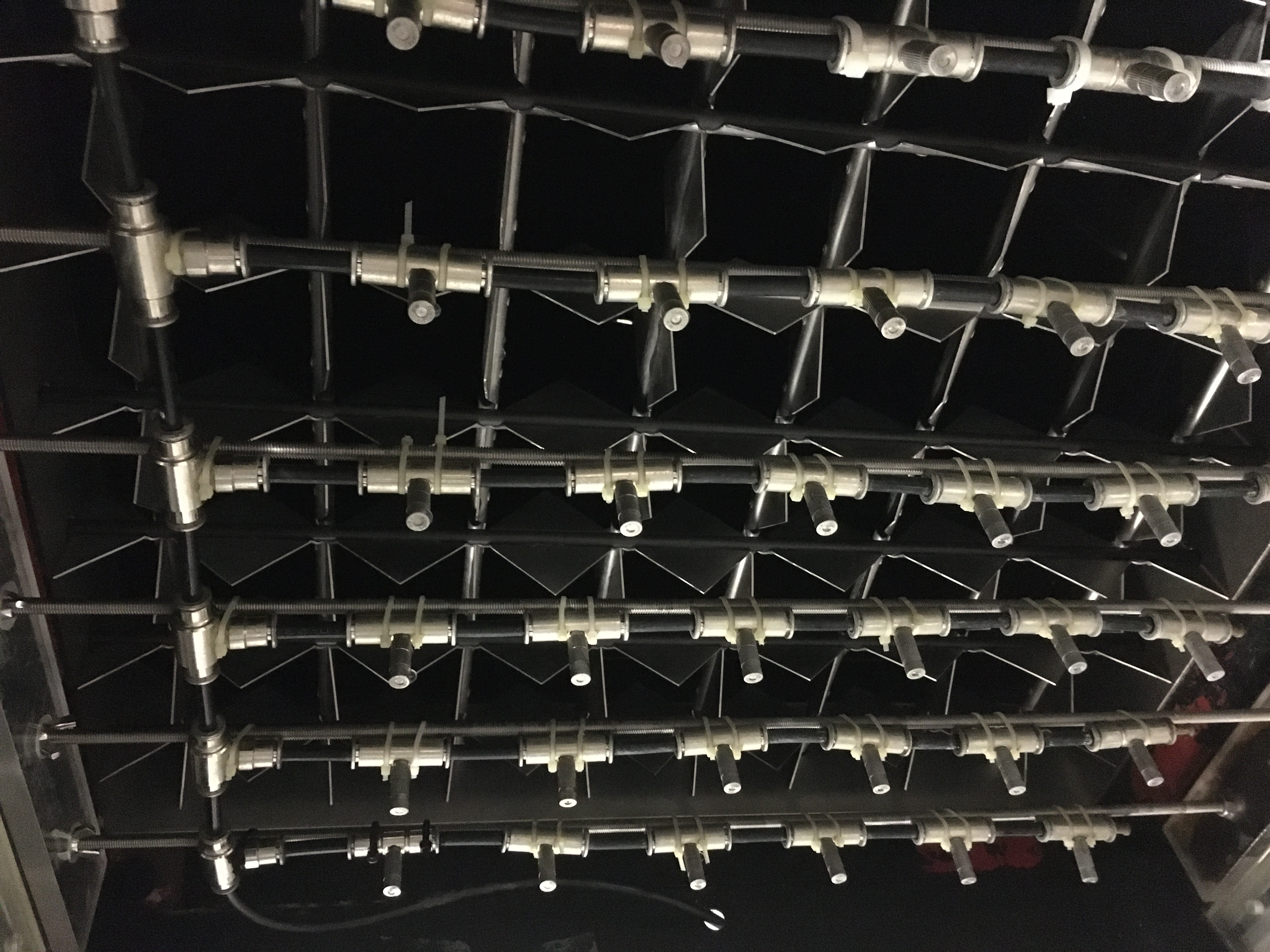}
        \caption{}
        \label{fig:grille}
    \end{subfigure}
    ~
    \begin{subfigure}[b]{0.58\textwidth}
        \centering \includegraphics[width=\textwidth]{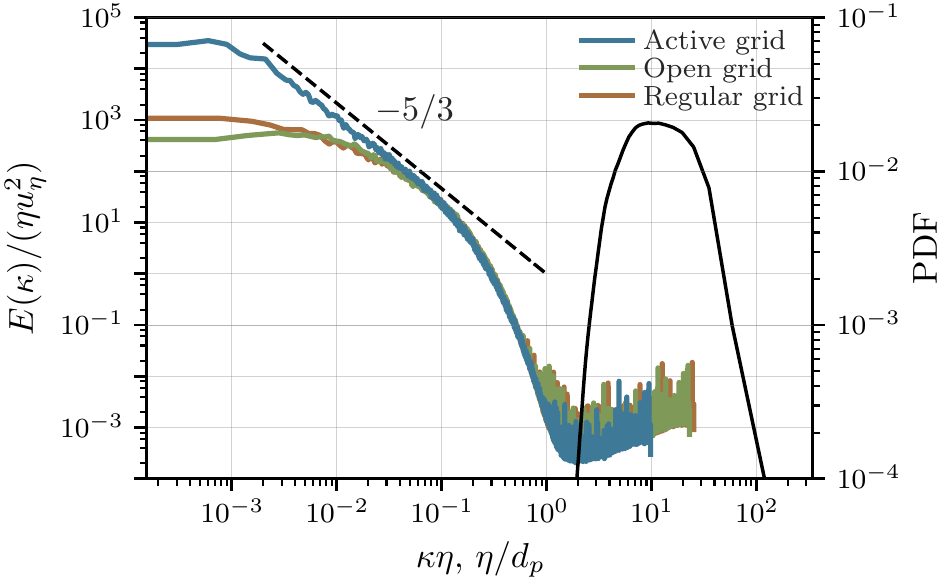}
        \caption{}
        \label{fig:wt_spectre}
    \end{subfigure}
    \caption{(a) Sketch of the wind tunnel with the PDPA measurement system. (b) Picture of the droplet injection system and, behind it, of the active grid in open grid mode. (c) Power spectral density of the longitudinal velocity from hot-wire records normalised by the Kolmogorov scale for an inlet velocity around \SI{4}{m/s}. The dashed line presents a Kolmogorov $-5/3$ power law scaling, as reference. The average inertial particle diameter distribution, normalised by the Kolmogorov scale is shown on the right axis. Note that it is plotted against $\eta/d_p$.}
    \label{fig:experimental_setup}
\end{figure}

Experiments were conducted in the Lespinard wind tunnel, a closed-circuit wind tunnel at LEGI  (Laboratoire des Ecoulements Géophysiques et Industriels), Grenoble, France. 
The test section is \SI{4}{\m} long with a cross section of \SI{0.75x0.75}{\m}.  A sketch of the facility is shown in the panel of Figure~\ref{fig:WT}.
The turbulence is generated with two different grids: a static (regular) and an active grid.
The regular grid (RG) is a passive grid composed by 7 horizontal and 7 vertical round bars forming a square mesh with a mesh size of \SI{10.5}{\cm}. 
The active grid is composed by 16 rotating axes (eight horizontal and eight vertical) mounted with co-planar square blades and a mesh size of \SI{9}{\cm}, (see \cite{Obligado2011, Mora2019energycascade} for further details about the active grid). Each axis is driven by a motor whose rotation rate and direction can be controlled independently. Two protocols were used with the active grid. In the active grid (AG) protocol (also referred to as ``triple-random'' in the literature \cite{JohanssonMakita1991, Mydlarski2017}), the blades move with random speed and direction, both changing randomly in time, with a certain time scale provided in the protocol. For the open-grid protocol (OG), each axis remains completely static with the grid fully open, minimising blockage. These two protocols have been shown to create a large range of turbulent conditions, from $Re_\lambda \sim 30 $ for OG to above 800 for AG (\cite{Mora2019energycascade, Obligado2020}).

The turbulent intensity $u'/U_\infty$ obtained for OG is in the same range as for RG, $\approx 2-3\%$.
The turbulent intensity created by the AG is much larger, just below $15\%$.
However, some significant differences exist between RG and OG turbulence: the bar width of the regular grid is twice that of the open grid (2 cm  vs. 1 cm) and the open grid has a 3D structure due to the square blades (see Figure~\ref{fig:grille} for an illustration of the OG).
This implies significant differences in the integral length scale $\mathcal{L}$ of the turbulence, $\approx 6~cm$ for RG versus $\approx 3~cm$ for RG. 
These various grid configurations allowed us to explore different Taylor-scale Reynolds numbers $Re_\lambda$, from 34 to 513 at a fixed freestream velocity. Additionally, our experimental setup  allowed for the study of particles at similar values of $u'/U_\infty$ and $Re_\lambda$, but different $\mathcal{L}$ (with OG versus RG). Matching the AG Reynolds number with the passive grids was not possible as it would require high wind tunnel velocities in the RG/OG cases, which would limit the measurements of the settling velocity due to low resolution.

Hot-wire anemometry (HWA) measurements were taken to characterise the single-phase turbulence  (\cite{Mora2019energycascade}). %in absence of particles. 
A constant temperature anemometer (Streamline, Dantec Inc. Skovlunde, Denmark) was used with a 55P01 hot-wire probe (\SI{5}{\um} in diameter, \SI{1.25}{mm} in length). %https://www.dantecdynamics.com/components/hot-wire-and-hot-film-probes/single-sensor-probes/gold-plated-wire/
The hot-wire was aligned with the centreline of the tunnel, (\SI{3}{\m} downstream the turbulence generation system). 
Additional measurements were carried out near the wall of the wind tunnel to check the homogeneity of the turbulence characteristics.
Velocity time series were recorded for \SI{180}{s} with a sampling frequency $F_s$ of \SI{50}{kHz}. This sampling frequency provides adequate resolution down to the Kolmogorov length scale $\eta$.

The background flow was also characterised with a Cobra Probe: a multi-hole pressure probe which is able to capture three velocity components. 
This multi-hole pitot tube probe (Series 100 Cobra Probe, Turbulent Flow Instrument TFI, Melbourne, Australia) was used to characterise possible contributions of the non-streamwise velocity components to the average value. 
The acquisition time of the measurements was set to \SI{180}{s} with a data rate of \SI{1250}{Hz} (the maximum attainable). As the turbulence scales may reach beyond this frequency, and may not be resolved due to the finite size of the probe, which has a sensing area of \SI{4}{mm^2} (\cite{ Mora2019energycascade, Obligado2022}), these measurements are used only to compute the mean and rms values of the 3D velocity vector. To estimate the small angle present between the probe head and the direction of the mean flow, measurements were collected in  laminar flow conditions (i.e, without any grid in the test section), to estimate the misalignment angle between the Cobra head and the streamwise direction.\\ 

Single-point turbulence statistics were calculated for each flow condition. The turbulent Reynolds number based on the Taylor microscale is defined as  $Re_\lambda = u' \lambda/\nu$ where $u'$ is the standard deviation of the streamwise velocity component, $\nu$ the kinematic viscosity of the flow and $\lambda$ the Taylor microscale.
The Taylor microscale was computed from the turbulent dissipation rate $\varepsilon$ with $\lambda = \sqrt{15\nu u'^2/\varepsilon}$, extracted as  $\varepsilon = \int 15 \nu  \kappa^2 E(\kappa) d\kappa$ where $E(\kappa)$ is the energy spectrum along the wavenumber $\kappa$. % and $k$ is . say what is k
The small scales of the turbulent flow are characterised by the Kolmogorov length, time and velocity scales: $\eta = (\nu^3/\varepsilon)^{1/4}$, $\tau_\eta = (\nu/\varepsilon)^{1/2}$ and  $u_\eta = (\nu\varepsilon)^{1/4}$. 
Different methods were used to estimate the integral length scale.
$\mathcal{L}$ was first computed by direct integration of the autocorrelation function until the first zero-crossing $\mathcal{L}_{a} = \int_0 ^{\rho_\delta} Ruu(\rho) d\rho$  and until the smallest value of $\rho$ for which $Ruu(\rho_\delta)=1/\exp$ (\cite{Puga2017, Mora2019energycascade}). 
The integral length scale was also estimated from a Vorono\"i analysis of the longitudinal fluctuating velocity zero-crossings $\mathcal{L}_{voro}$, following the method recently proposed in (\cite{Mora2020integralscale}), where an extrapolation of the 1/4 scaling law was performed when needed.
The latter is particularly relevant for the active grid mode, where the value of $R_{uu}$ has been found, in some cases, to not cross zero (\cite{Puga2017}).

\begin{figure}[hbtp!]
	\centering
	\includegraphics[width=\textwidth]{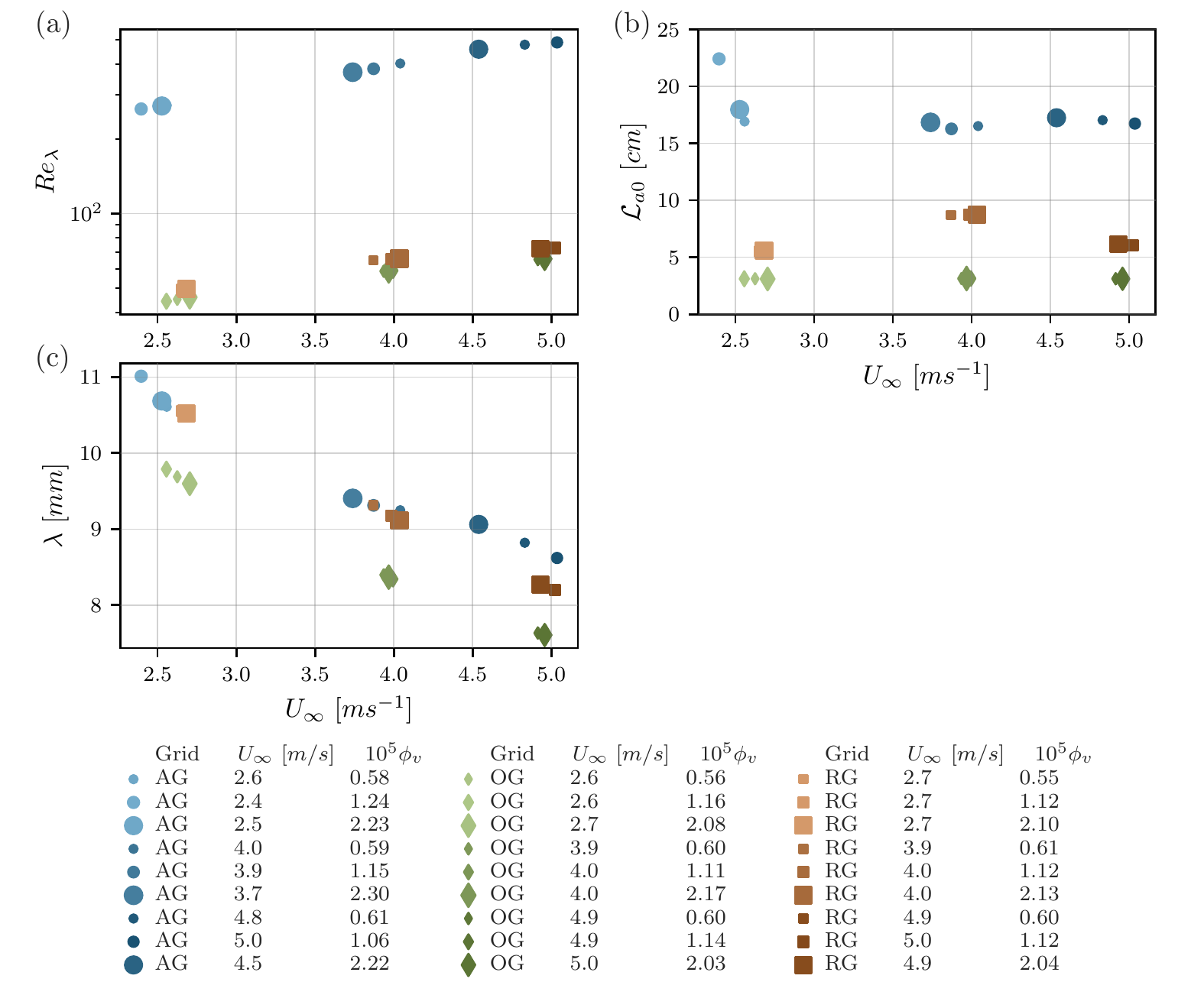}	
	\caption{(a) Taylor Reynolds number $Re_\lambda$, (b)  integral length scale from the integration of the autocorrelation to the first zero-crossing, (c) Taylor microscale $\lambda$. All plotted versus the mean streamwise velocity obtained from hot-wire measurements. The different symbols ($\medblacksquare$), ($\bullet$) and ($\smallblackdiamond$) represent the regular, active and open grid respectively. 
	The size of the symbol is proportional to the volume fraction and darker colours correspond to higher mean velocities. 
	}
	\label{fig:turbulence}
\end{figure}

Table \ref{tab:turbulence_param} summarises the flow parameters for all experimental conditions studied. 
The right panel of Figure \ref{fig:wt_spectre} shows the power spectral density of the streamwise velocity computed from hot-wire time signals at the measurement location ($x \approx 3$ m for all cases).
The three spectra depicted in the figure were obtained from the three different grid configurations, all of them with an inlet velocity of approximately \SI{4}{m/s}.
The power spectral density was normalised by the Kolmogorov length and velocity scales $\eta$ and $u_\eta$.
As expected, the turbulent flow generated by the active grid exhibits a considerably wider inertial range.
On the right of the figure, the mean droplet diameter distribution is displayed. 
The diameter distribution, discussed in the next section, was normalised by the smallest Kolmogorov scale among all conditions (i.e. the Kolmogorov scale of the active grid turbulent flow).
It can be observed that the distribution is polydisperse and particles are always much smaller than the Kolmogorov scale of the turbulence.
Figure \ref{fig:turbulence} shows the Taylor Reynolds number $Re_\lambda$ and the Taylor microscale $\lambda$ for different wind tunnel velocities \SI{3}{m} downstream (at approximately $x/M \approx 30$). 

\begin{table}[ht]
	\centering
	\begin{tabular}{lccc}
		Parameters  & AG & OG & RG \\
		\hline
		$U_\infty ~ (ms^{-1})$ & 2.6 - 5.0  & 2.6 - 5.0 & 2.6 - 5.0 \\
		$Re_\lambda$ & 268 - 513 & 34 - 55 & 49 - 68 \\
		$u'/U_\infty ~(\%)$ & 13.2 - 14.9 & 1.9 - 2.1 & 2.5 - 2.7\\
		$10^3 \times \varepsilon ~ (m^2s^{-3})$ &  140.1 - 1251.4 & 6.9 - 26.8 & 9.9 - 59.5 \\
		$\eta ~(\mu m)$ & 230 - 406  & 634 - 868 & 511 - 792 \\
		$\tau_\eta ~(ms)$ & 3.5 - 11.0 & 26.7 - 50.2 & 17.4 - 41.9 \\
		$\lambda ~(cm)$ & 1.02 - 1.29 & 0.92 - 1.16 & 0.83 - 1.09  \\
		$\mathcal{L}_{a0} ~ (cm)$ & 16.3 - 22.4 & 3.0 - 3.1 & 5.5 - 8.7 \\
		$\mathcal{L}_{a\delta} ~ (cm)$ & 8.5 - 9.6 & 1.8 - 1.9 & 2.2 - 2.4 \\
		$\mathcal{L}_{voro} ~ (cm)$ & 14.0 - 24.0 & 2.3 - 2.8 & 3.7 - 4.5 \\
		\hline	
	\end{tabular}
\caption{Turbulence parameters for the carrier phase, sorted by grid category computed from hot-wire anemometry measurements 3 meters downstream of the grid. $U_\infty$ is the freestream velocity, $u'$ the rms of the streamwise velocity fluctuations, $Re_\lambda = u' \lambda/\nu$ the Taylor-Reynolds number and $\varepsilon = 15\nu u'^2/\lambda^2$ the turbulent energy dissipation rate. $\eta = (\nu^3/\varepsilon)^{1/4}$ and $\tau_\eta = (\nu/\varepsilon)^{1/2}$ are the Kolmogorov length and time scales. $\lambda = \sqrt{15\nu u'^2/\varepsilon}$ and $\mathcal{L}$ are the Taylor microscale and the integral length scale, respectively, where three different methods are used to compute $\mathcal{L}$.} 
\label{tab:turbulence_param}
\end{table}

\subsection{Particle Injection}
Water droplets were injected in the wind tunnel by means of a rack of 18 or 36 injectors distributed uniformly across the cross-section. 
The outlet diameter of the injectors is  of \SI{0.4}{mm}, and atomization is produced by high-pressure at 100 bars.
The particle volume fraction $\phi_v = F_{water}/F_{air}$ describes the ratio between the water and air volumetric flow rates.
The volume fraction in the experiments $\phi_v$ was varied in the range $\phi_v \in [0.5 \times 10^{-5}, 2.0 \times 10^{-5}]$.
18 or 36 injectors were used depending on the experimental conditions, as low volume fractions could not be reached with 36 injectors. 
The resulting inertial water droplets have a polydisperse size distribution with a $D_{max}$ and $D_{32}$ of $\approx30~\mu m$ and $\approx 65~\mu m$,  respectively (\cite{Sumbekova2017}), as shown in Figure \ref{fig:wt_spectre}, with $D_{32}$ the Sauter mean diameter.
The droplet Reynolds numbers $Re_p$ are smaller than one.
For each grid mode, three different volume fraction were tested, with three different freestream velocities ($U_\infty =  2.6,~  4.0,~  5.0$ m/s).
This results in 27 different experimental conditions.\\

Measurements were collected with a Phase Doppler Analyzer (PDPA) (\cite{Bachalo1984}). The PDPA (PDI-200MD, Artium Technologies, Palo Alto, CA. USA) is composed of a transmitter and a receiver positioned at opposite sides of the wind tunnel. The transmitter emits two solid-state lasers, green at \SI{532}{\nm} wavelength and blue at \SI{473}{\nm} wavelength. 
Both lasers are split into two beams of equal intensity and one of these is shifted in frequency by 40 MHz, so that when they overlap in space they form an interference pattern. The \SI{532}{\nm} beam enables us to take the particle's vertical velocity and diameter simultaneously.  
The second beam %(473 nm wavelength) 
is oriented to measure the horizontal velocity.
The PDPA measurements were non-coincident, i.e. horizontal and vertical velocities were taken independently, since recording only coincident data points can significantly reduce the validation rate. The particle's horizontal velocity $\langle U \rangle$ is assumed to be very close to the unladen incoming velocity $\langle U \rangle \approx U_\infty$. Contrary to the study of \cite{Mora2021settling}  in the same facility, the transmitter and the receiver had a smaller focal length of \SI{500}{\mm}.
This enable us to measure the particle vertical velocity with better resolution. 
The vertical and streamwise velocity components were recorded with a resolution of \SI{1}{mm/s}.
The PDPA configuration allow us to detect particles with diameters ranging from \SI{1.5}{\um} to \SI{150}{\um}. We verified that all velocity distribution were Gaussian, as expected under HIT conditions (see appendix \ref{A2}).
The measurement volume was positioned \SI{3}{\m} downstream of the droplet injection (at approximately the same streamwise distance as the hot-wire and Cobra measurements).
In order to quantify the effect of recirculation currents, data were collected on the centreline of the wind tunnel and at a off-centre location, 10 cm from the wind tunnel wall.
For each set of experimental conditions, at least $5 \times 10^5$ samples were collected. Depending on the water flow rate and the wind tunnel inlet velocity, the measurement sampling rate varied from 20 Hz to 4800 Hz with an average of 1030 Hz and 580 Hz for the streamwise and vertical velocities, respectively.
\newline

\subsection{Angle correction}
As the settling velocity is only a small fraction of the  the particle velocity, any slight misalignment of the PDPA with the  vertical axis ($y$) would result in a large error on the measurements of this important variable. To correct the optical alignment bias, the misalignment angle $\beta$ was computed from very small ($d_p < 4~\mu m$) olive oil droplets measurements, as described in \cite{Mora2021settling}. Olive oil generators produce a monodisperese droplet distribution  ($\langle d_p \rangle  \approx 3 ~ \mu m$), that behave as tracers. Using the empirical formula from Schiller \& Nauman (\cite{Bubblesdrops}) for the settling velocity of particles, and assuming that the mean centreline velocity is purely streamwise, the misalignment between the PDPA and gravity was estimated. Data from the alignment bias correction is given in appendix \ref{olive_oil}.
The angle $\beta$ was determined to be $\beta = 1.5 \degree \pm 0.3 \degree $. The vertical velocity measurements were then corrected subtracting the $V_\beta$ misalignment bias (proportional to the streamwise velocity and the sine of the misalignment angle).

\section{Results}

\subsection{Settling velocity of inertial particles as a function of size.} 
\begin{figure}[htp]
	\centering 
	\includegraphics[width=\textwidth]{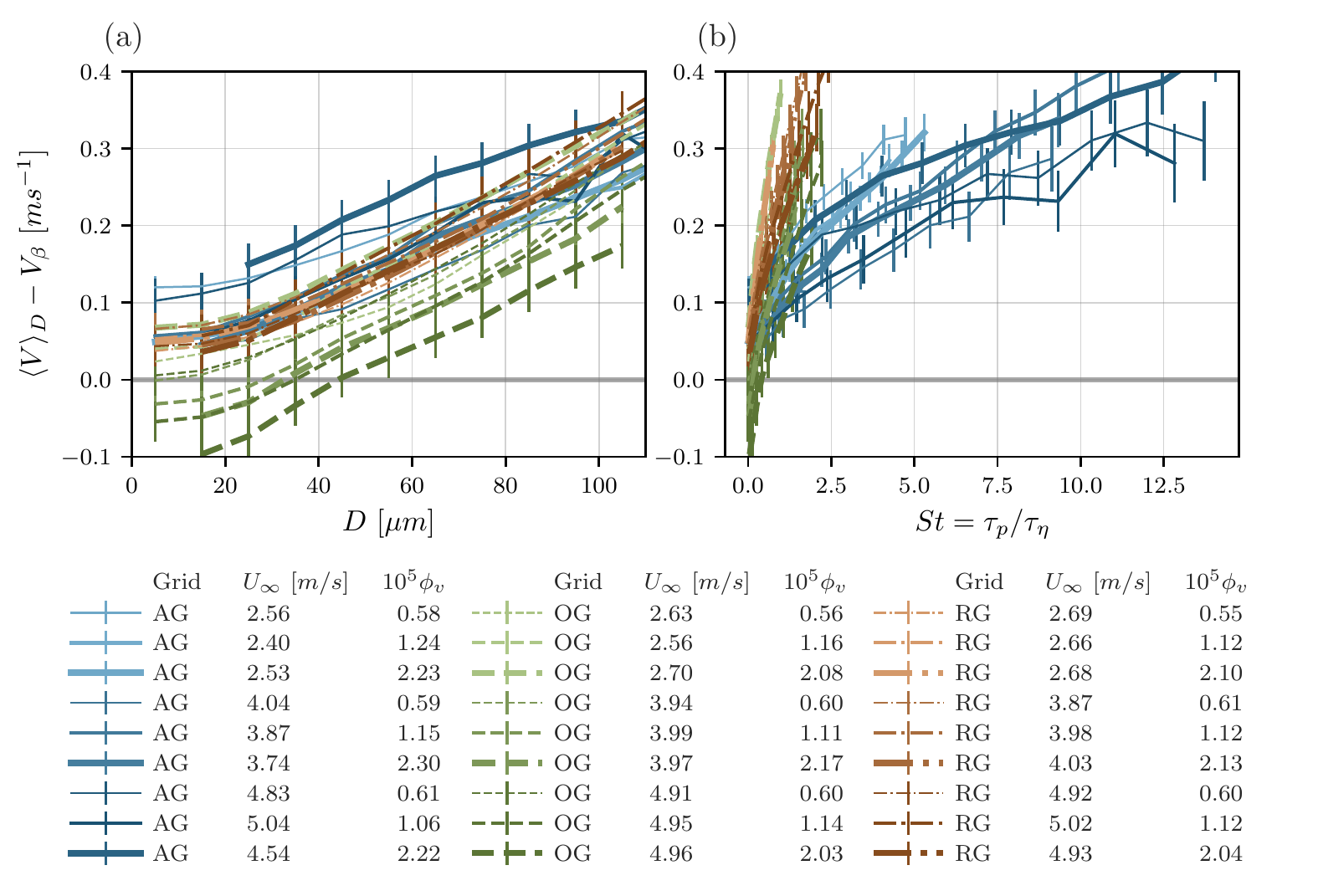}
	\caption{Corrected particle vertical velocity $\langle V \rangle_D - V_{\beta}$ averaged over bins of  \SI{10}{\um} against the diameter (a) and the Stokes number (b). The data from the active grid (AG) are in solid lines, the open grid (OG) in dashed line and the regular grid (RG) in dash-dotted line. The error bars show the estimation of the error in the velocity measurements. Darker colours correspond to higher mean velocities $U_\infty$ and the line width is proportional to the volume fraction.}
	\label{fig:VvsD}
\end{figure}

Figure \ref{fig:VvsD} presents the corrected averaged settling velocity $\langle V \rangle_D - V_{\beta}$ against the diameter $D$ and the Stokes number $St$. Vertical velocity is defined as positive when downwards. In all figures, we averaged the settling velocity in \SI{10}{\um} bins, from 0 to \SI{150}{\um}.

For each experimental conditions, as expected, the velocity measurements show that, on average,  larger particles have higher settling velocity. 

\subsection{Non-zero mean vertical flow in the limit of very small diameter}
In \cite{Maxey1987}, Maxey showed that in the limit of zero particle inertia ($St << 1$), the mean particle settling velocity $\langle V(t) \rangle$ is the sum of the Eulerian mean fluid velocity and the still-fluid settling velocity,
\begin{equation}
    \langle V(t) \rangle |_{St << 1} = \langle U_y \rangle + V_T.
\end{equation}
In the limit of no particle inertia, the particle relaxation time $\tau_p$ tends to zero, and therefore $V_T$ (which can be computed as $V_T=g\tau_p$) also tends to zero. Consequently, in the zero-inertia limit and for very dilute conditions, particles should behave as tracers and follow the fluid streamlines. Assuming that the air flow has no mean motion in the vertical direction in the centreline, the mean corrected vertical particle velocity $\langle V \rangle_D - V_{\beta}$ should tend to zero for small diameters.

However, experimental data shown in Figure \ref{fig:VvsD} present an offset velocity when the diameter tends to zero.
This offset velocity for very small particle was already encountered in this facility (\cite{Sumbekova2016phd, Mora2021settling}) and suggests a vertical component due to secondary motion in the air in the wind tunnel, $\langle U_y \rangle \neq 0$. 
A mean gas velocity in the vertical direction could be due to two different physical phenomena. First, as discussed previously, confinement effects (that would be different for each type of the grid) can be responsible for secondary recirculation motion inside the tunnel. Second, the injection of droplets could modify the background flow, since falling droplets may entrain gas in their fall. Entrainment in the wake of falling particles might induced a downward mean gas flow, with a velocity that should be proportional to the dispersed-phase volume fraction (\cite{Alipchenkov2009, Sumbekova2016phd}). 
To compensate the downward gas secondary motion near the centreline of the wind tunnel, an upwards flow in the gas near the walls should be present (and viceversa for upwards gas velocity at the centreline).

Other studies have encountered similar difficulties due to recirculating secondary motions when measuring particle settling velocity (\cite{Wang2018, Akutina2020}).
\cite{Akutina2020} corrected for this bias by subtracting the local mean fluid velocity measurements from the instantaneous vertical velocity of the particle (available in the point-particle simulations). 

\begin{figure}[t!]
	\centering 
	\includegraphics[width=\textwidth]{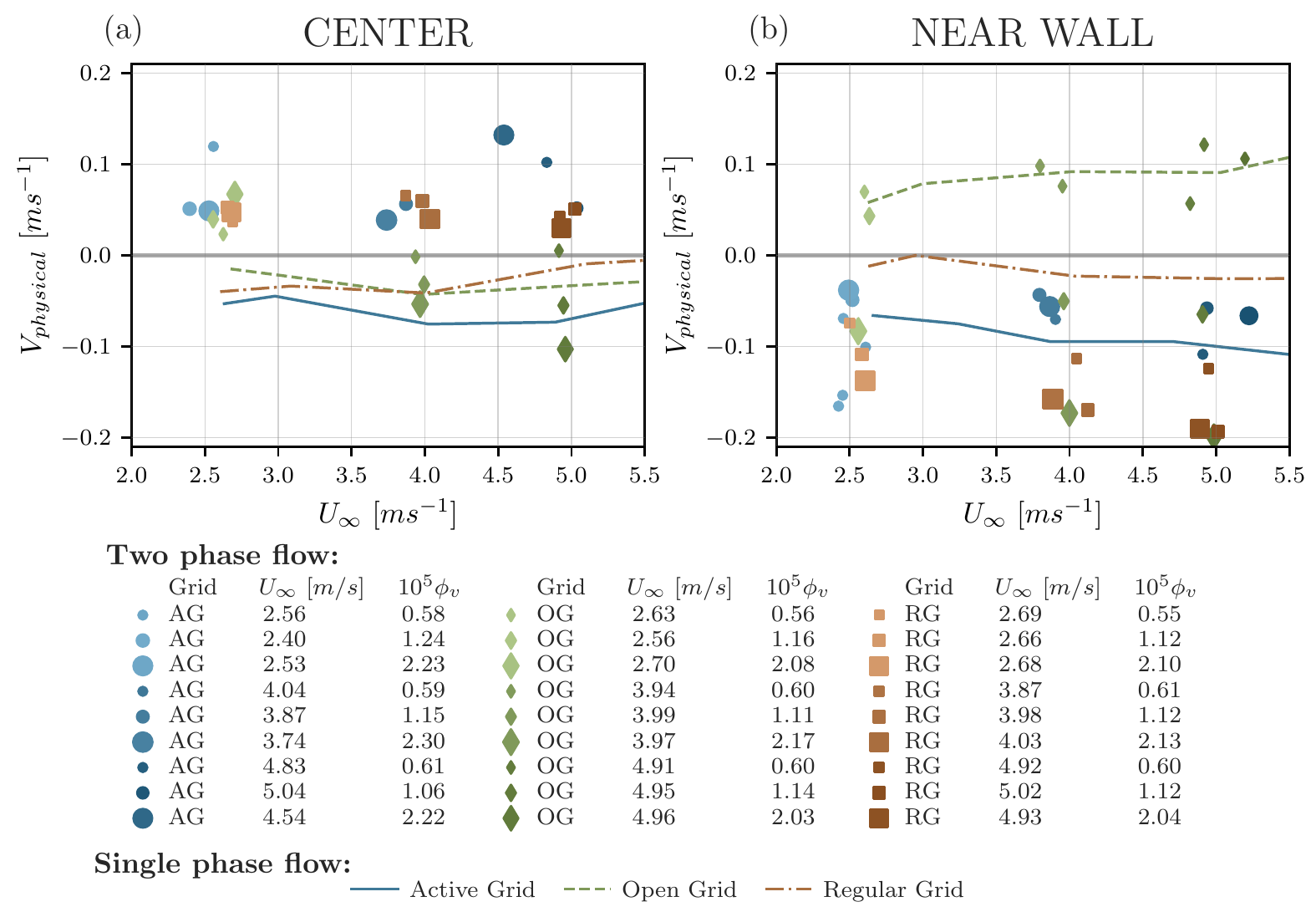}
	\caption{Average settling velocity of the particles for the smallest diameter class, (a) at the centre, and (b) near the wall of the wind tunnel.
	The different symbols represent the regular ($\medblacksquare$), active ($\bullet$) and open grid ($\smallblackdiamond$). The size of the symbols is proportional to the volume fraction and a darker colour correspond to a higher mean velocity. Carrier-phase vertical velocity measurements with the Cobra probe are presented at the two locations with coloured lines.
	Similar to Figure \ref{fig:VvsD}, active grid (AG), open grid (OG) and regular grid (RG) are in solid line, dashed line and dash-dotted line respectively.
	}
	\label{fig:Vphysical}
\end{figure}

We estimated the existence and strength of recirculating secondary motion in the wind tunnel by taking PDPA measurements in the centre and close to the wall of the wind tunnel. We quantified the carrier-phase vertical velocity using the mean settling velocity of the smallest particles with enough statistical convergence. This parameter is referred to as $V_{physical}$. Figure \ref{fig:Vphysical} shows $V_{physical}$, measured  
in the centre (left panel) and near the wind tunnel wall (right panel).

Figure \ref{fig:Vphysical} shows downward motion ($V_{physical} > 0$) at the centre and upward motion ($V_{physical} < 0$) near the wind tunnel sidewall, in most cases.
A different behaviour is observed for the open grid (star symbols), with opposite direction of secondary motion, for some volume fractions.
There is no clear trend for $V_{physical}$ with the volume fraction, probably because the range of volume fraction investigated is not significant for this phenomenon.

We also observed recirculating secondary motions in the single-phase flow measured with the Cobra probe.
Lines in Figure \ref{fig:Vphysical} show the mean single-phase vertical velocity for the three turbulence conditions, against the mean streamwise velocity.
Measurements with the Cobra probe provide evidence that there are weak secondary flows in the wind tunnel, even in the absence of particles.
Moreover, these secondary flows are dependent on the turbulence generation mechanism, as the open grid (dashed line) causes an opposite sense of motion than the active or regular grids. 
Surprisingly, single-phase measurements confirm the same trends as the particle velocity measurements. At the most dilute case (i.e. for the lowest volume fraction, the vertical velocity of the secondary motion is the same order of magnitude in the single- and two-phase flows: \SI{0.1}{m/s}).

To conclude, measurements in both laden and unladen flows show the existence of downward motion in the centre and upward motion near the sidewalls (with the active and regular grids, with the opposite sense of motion for the open grid). 
To the best of the authors' knowledge, this constitutes the first experimental evidence on the existence of $V_{physical}$ as a quantification of the carrier-phase vertical velocity in wind tunnel experiments.
From now on, $V_{physical}$ and $V_\beta$ are subtracted from the measurements of vertical velocity, $\langle V \rangle_D - V_{\beta} - V_{physical}$, to quantify settling velocity enhancement and/or hindering (corrected from these two experimental biases).

\subsection{Influence of the carrier flow turbulent Reynolds number on the particle settling velocity}\label{res}

To quantify modifications of the settling velocity, we subtract the particle terminal speed in a stagnant fluid $V_T$ from the vertical velocity. We define this difference as $\Delta V$, where positive values imply settling velocity enhancement and negative correspond to hindering. The value of $V_T$ is estimated using the Schiller \& Nauman empirical formula.
$\Delta V$ is usually normalized by the rms of the carrier-phase fluctuations, $u'$, or by the particle terminal velocity, $V_T$.  
Normalising $\Delta V$ by $u'$ was first proposed by \cite{wang_maxey_1993}, and \cite{Yang1998} confirmed  $u'$ is a better velocity scale than $u_\eta$ to express the settling velocity enhancement. It has been widely used in other studies (\cite{Rosa2016, Huck2018}). Consequently, $\Delta V$ is normalized by $u'$, although this non-dimensionalisation of $\Delta V$ is still under scrutiny.

\begin{figure}[t!]
	\centering 
	\includegraphics[width=\textwidth]{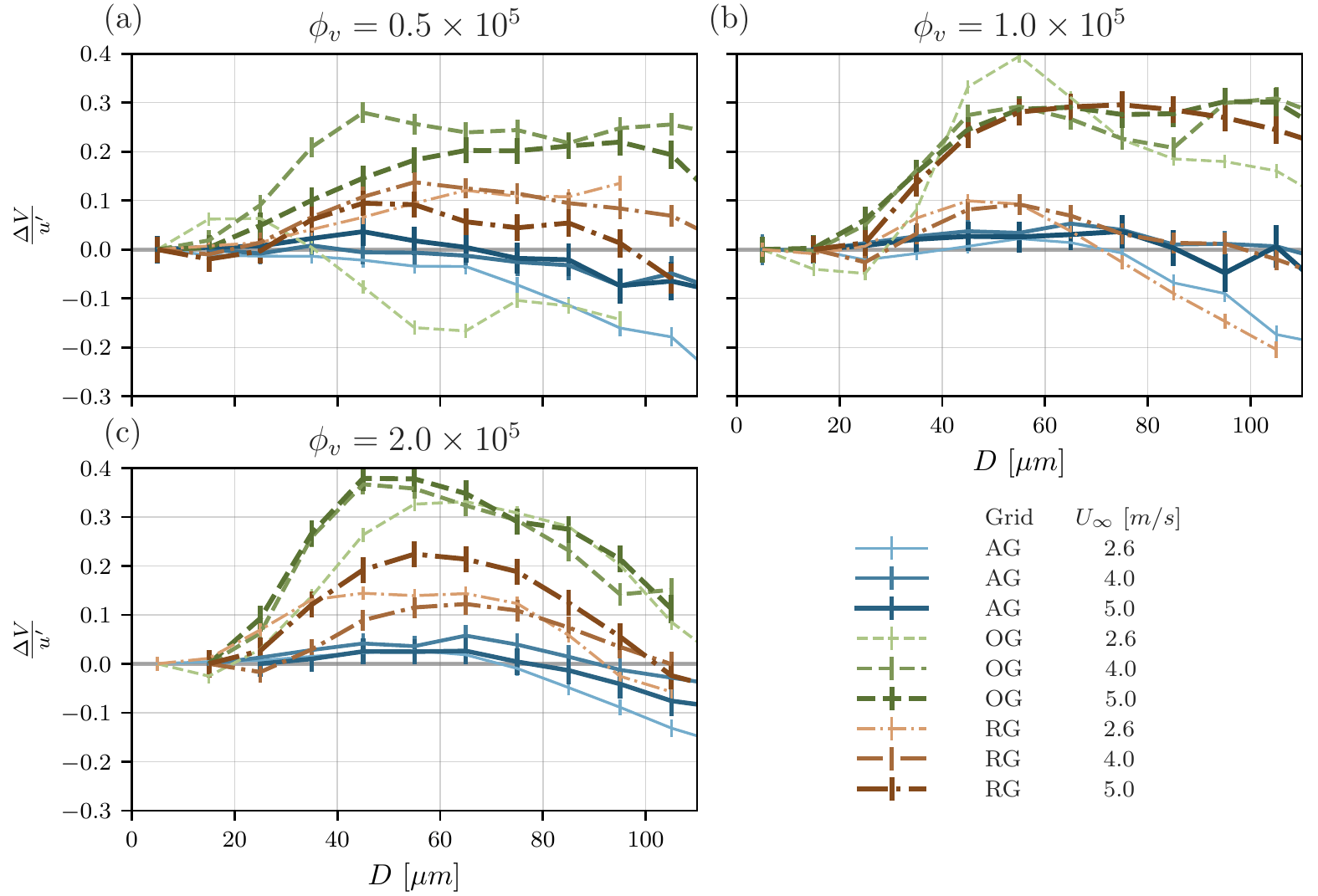}
	\caption{Particle velocity over the carrier phase fluctuations $\Delta V/u' = (\langle V \rangle_D - V_{\beta} - V_{physical} - V_T)/u'$ against the particles diameter $D$ for a volume fraction of $0.5\times10^{-5}$ (a), $1.0\times10^{-5}$ (b) and $2.0\times10^{-5}$ (c).
   	The data from the active grid (AG) are in solid lines, the open grid (OG) in dashed line and the regular grid (RG) in dash-dotted lines. 
   	The errorbars show the estimation of the error in the velocity measurements induced by the determination of the misalignement angle.
    A darker color correspond to a higher mean velocity $U_\infty$.}
	\label{fig:DV_vs_D}
\end{figure}

Figure \ref{fig:DV_vs_D} shows the normalised velocity difference $\Delta V/u'$ against particle diameter. 
All the measurements were taken at the same location, at the centreline of the wind tunnel.
All the curves show the same trend: the settling velocity is enhanced for small particles, and this enhancement reaches a maximum, $\max(\Delta V /u')$.
After the maximum, the settling velocity enhancement decreases until it reaches a point where it is negative, that is, particle settling is hindered by turbulence. 
For very large particles (not attainable with our injection system), $\Delta V /u'$ would eventually become zero as they follow ballistic trajectories, unimpeded by turbulence. 

Particle settling velocity tends to depend on the turbulence characteristics, that is, in this study, it depends on the type of grid used in the experiments. Series taken with the open-grid configuration show a higher enhancement for all volumes fractions (green dashed line).
On the contrary, active-grid turbulence (in blue solid lines) causes mostly hindered settling, with enhancement  present only for a small range of diameters.
Finally, measurements taken with the regular grid (red dash-dotted lines) show an intermediate behaviour between the two other grid configurations.\\

A combination between the Rouse and Stokes numbers, $RoSt$, has already been proven to be an interesting scaling (\cite{ghosh2005}), as it was shown  in several studies to collapse the data better (\cite{Good2014, Petersen2019, Mora2021settling, Yang2021}).
The Rouse-Stokes number can be expressed as a ratio between a characteristic length of the particle $L_p$ and a characteristic length  of the flow. Using the Kolmogorov time scale in the Stokes number and $u'$ in the Rouse number, the Taylor microscale appears to be the characteristic length scale of the flow:

\begin{equation}
    RoSt = \frac{\tau_p}{\tau_\eta}\frac{V_T}{u'} = \sqrt{15} \frac{V_T \tau_p}{\lambda} = \sqrt{15} \frac{L_p}{\lambda} \quad \text{with} \quad L_p = V_T\tau_p \quad \text{as} \quad \lambda = \sqrt{15}\tau_\eta u'
\end{equation}

\begin{figure}[t!]
	\centering 
	\includegraphics[width=\textwidth]{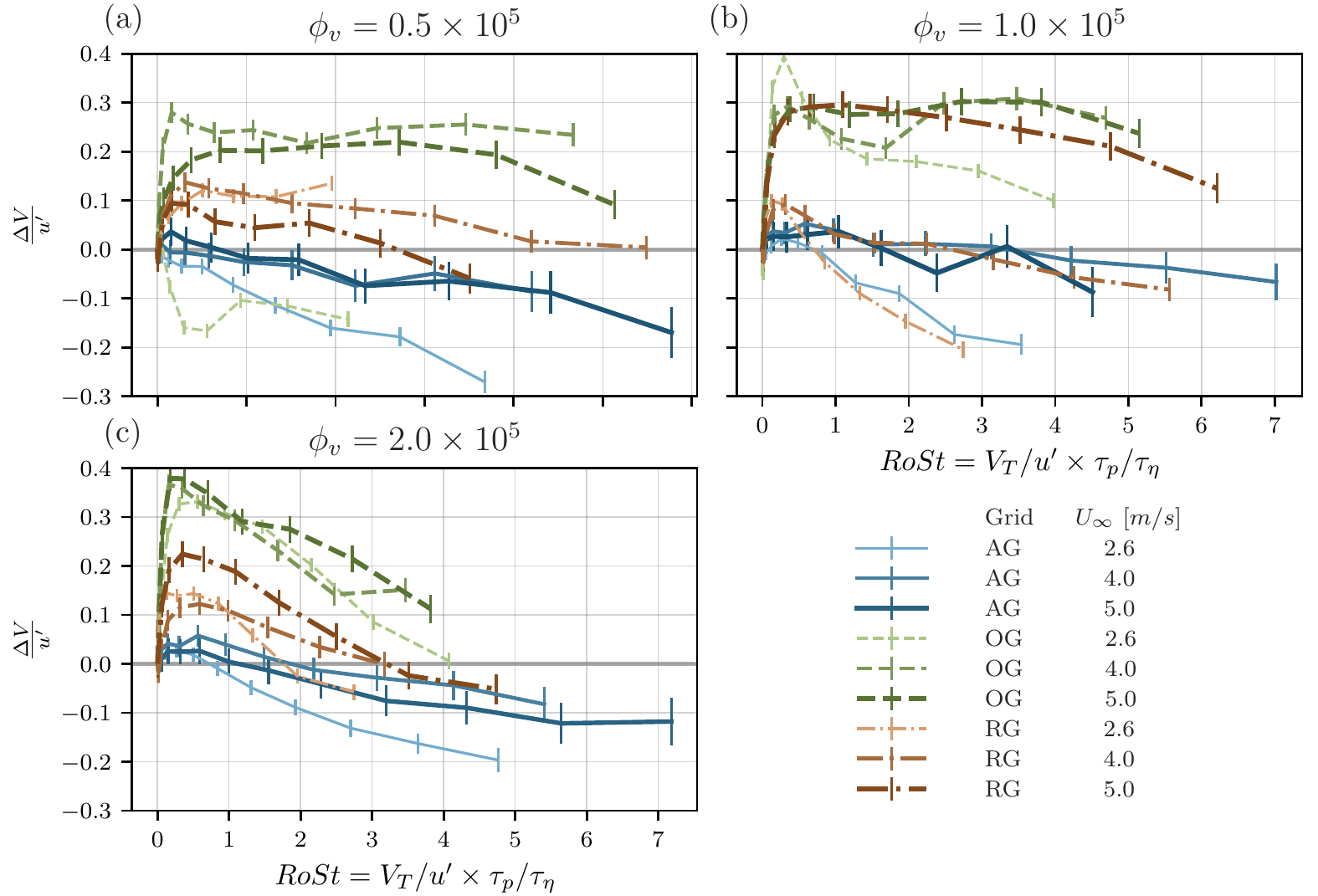}
	\caption{Enhancement of the particle velocity, normalised by the turbulent rms velocity, $\Delta V/u'$, against the Rouse-Stokes number.  (a) $\phi=0.5\times10^{-5}$, (b) $\phi=1.0\times10^{-5}$ and (c) $\phi=2.0\times10^{-5}$.
	Lines follow the legend of Figure \ref{fig:DV_vs_D}. 
    }
	\label{fig:DV_vs_RoSt}
\end{figure}

In Figure \ref{fig:DV_vs_RoSt}, we present $\Delta V/u'$ against the Rouse-Stokes number $RoSt$. Similar to Figure \ref{fig:DV_vs_D}, each panel presents data from a different value of volume fraction. 

The $RoSt$ number gives a better collapse of the position of maximum of enhancement than the Rouse number or Stokes number alone.
Figure \ref{fig:DV_vs_RoSt} indicates that enhancement of the settling velocity reaches a maximum for a Rouse-Stokes number around 0.6, which is consistent with previous findings. \cite{Yang2021} reported a maximum for a $RoSt$ around 0.72-1 in a Taylor Couette flow, whereas \cite{Petersen2019} presented a maximum of enhancement for $RoSt$ of order 0.1. Alternative scalings have been tested on our data, with the results provided for completion in appendix \ref{A1}. 
These measurements reveal that, for a fixed $Re_\lambda$, the enhancement increases with volume fraction, consistent with \cite{Aliseda2002, Monchaux2017}. 

We observe that the enhancement decreases with an increase in wind tunnel Reynolds number, as reported in \cite{Mora2021settling}. Indeed, the enhancement is much stronger for the open grid and the regular grid ($Re_\lambda \in [30-70]$) than for active grid measurements ($Re_\lambda \in [260-520]$) for all volume fractions.
While the study of \cite{Mora2021settling} obtain the same trend by adding data from the literature, in this study the entire range of Reynolds number was explored in the same facility. The opposite behaviour is observed at fixed turbulence intensity, that is, with the same grid turbulence generation system, we observe that, in most cases, the maximum of enhancement increases with the inlet velocity $U_\infty$.
This would suggest that the maximum of enhancement has a non-monotonic behaviour with the turbulent Reynolds number, as reported in (\cite{Mora2021settling}). A non-monotonic dependency of the degree of enhancement with Reynolds number has also been observed recently in \cite{Yang2021}.
This effect of $Re_\lambda$ on the maximum of enhancement confirms that the settling velocity modification is a multiscale phenomenon and one turbulent scale is not sufficient to characterise it (\cite{Tom2019}).

\subsection{Scaling of the maximum of enhancement}

As no theoretical consensus have been found on the settling velocity modification, empirical scalings are proposed. 
This study focuses on the value and location of maximum of enhancement $\max(\Delta V/u')$, and not on the critical  $RoSt$, where enhancement turns into hindering, as most cases with the passive grid did not reach the transition enhancement/hindering for high Rouse number, contrary to \cite{Mora2021settling}.
As the enhancement seems to increase with the wind tunnel Reynolds number velocity, $U_\infty$ is introduced through a global Reynolds number $Re_G = M U_\infty /\nu$, where $M$ is the mesh spacing in the turbulence-generating grid.
Several dimensionless parameters were tested to scale $\max(\Delta V/u')$: the global Reynolds number $Re_G$, the volume fraction $\phi_v$, the Taylor-scale Reynolds number $Re_\lambda$, a Reynolds number based on the integral length scale, and the $Ro$ or $St$ numbers corresponding to the maximum of enhancement.
The best scaling from the parameters above was found to be a combination of $Re_\lambda$, $Re_G$ and $\phi_v$.

\begin{figure}[h!]
	\centering 
	\includegraphics[width=\textwidth]{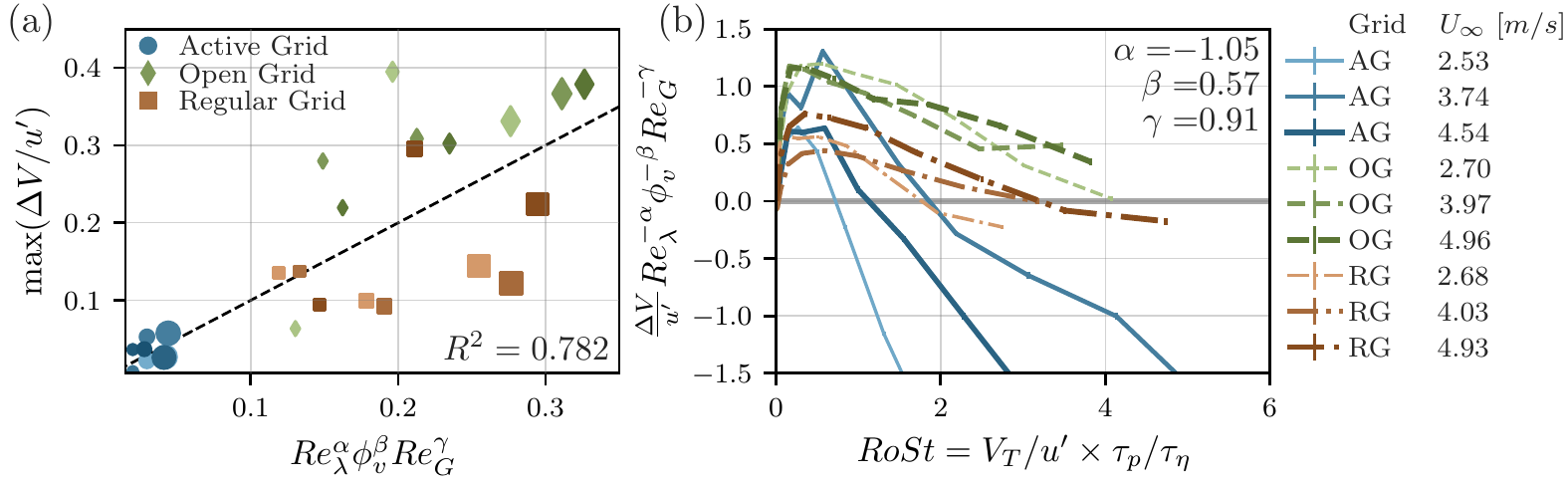}
	\caption{ Scaling of the settling velocity with $Re_\lambda$, $Re_G$ and $\phi_v$. (a) $\max(\Delta V/u')$ versus $Re_\lambda^\alpha \phi_v^\beta Re_G^\gamma$ with the fitted value of $\alpha$, $\beta$ and $\gamma$. (b) $\max(\Delta V/u')$ divided by the scaling versus the Rouse Stokes number.
	}
	\label{fig:scaling}
\end{figure}

Figure \ref{fig:scaling}(a) represents $\max(\Delta V/u')$ against $Re_\lambda^\alpha \phi_v^\beta Re_G^\gamma$, where $\alpha$, $\beta$ and $\gamma$ are best-fit exponents:
\begin{equation}
    \max(\Delta V/u') \sim Re_\lambda^\alpha \phi_v^\beta Re_G^\gamma 
\end{equation}

\noindent with $\alpha = -1.1$, $\beta = 0.6$ and $\gamma = 0.9$.
The values of $\alpha$, $\beta$ and $\gamma$ are consistent with previous observations: the maximum enhancement increases with inlet velocity and volume fraction but decreases with a global increase of $Re_\lambda$.

Figure \ref{fig:scaling} show the third panel of Figure \ref{fig:DV_vs_RoSt} with $\Delta V/u'$ divided by the power law scaling. A gap in data exists due to the jump in  Reynolds number between the active grid and the two passive grids (see Figure \ref{fig:turbulence}).
No measurements were taken for $Re_\lambda$ between 70 and 260, since the present experimental setup cannot reach those intermediate values. 

\section{Influence of large-scale structures}

Although the open and regular grids create very similar values of turbulent intensity, the settling speed of inertial particles in these two flows are very different.
Indeed, regular grid data (dash-dotted lines) is as different from open grid data as it is from active grid data (see Figures \ref{fig:DV_vs_D} and \ref{fig:DV_vs_RoSt}).
This discrepancy between regular and open-grid behaviours can be explained by the difference in integral length scales between these two turbulent flows (see table \ref{tab:turbulence_param} and Figure \ref{fig:turbulence}).

\begin{figure}[htb!]
	\centering 
	\includegraphics[width=0.6\textwidth]{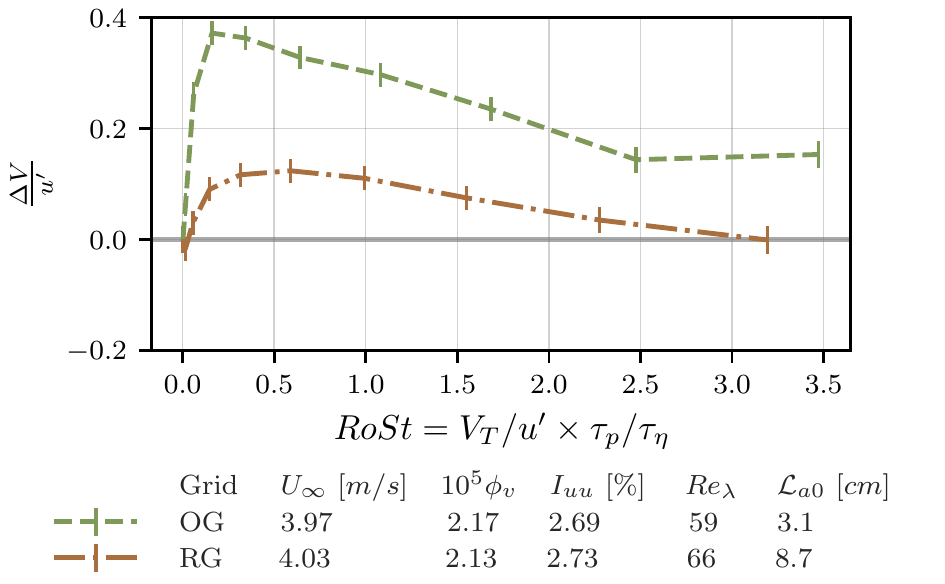}
	\caption{Open and regular grid data. Settling velocity difference over the carrier-phase fluctuations $(\Delta V)/u'$ against the Rouse-Stokes number, for a volume fraction of $2.0\times10^{-5}$ and an inlet velocity of \SI{4}{m/s}.}
	\label{fig:L}
\end{figure}

Figure \ref{fig:L} illustrates the settling velocity modification from two series with similar Reynolds numbers, turbulent intensities and volume fractions, but different integral length scales $\mathcal{L}_{a0}$. The figure is plotted against $RoSt$ but presents a similar trend when made with $Ro$ or $St$. It can be seen that the degree of settling enhancement is stronger for a smaller integral length scale and this behaviour is consistent for different volume fractions and wind tunnel Reynolds numbers. This suggests that the integral length scale and large-scale structures play a role in the settling velocity modification.
Figure \ref{fig:L} reveals that the integral length has an influence on the settling velocity modification for the entire range of RoSt number studied. 
However, \cite{Tom2019} shows that the flow scales contributing to the settling speed enhancement become larger as the Stokes number increases.

\section{Conclusion}

The settling velocity of sub-Kolmogorov inertial particles  in wind tunnel decaying turbulence is presented and analyzed.
Accurate settling velocity measurements were carefully collected and calibrated, by correcting different experimental sources of potential bias. First, a correction for PDPA misalignment angle is computed and applied. Second, secondary flows in the wind tunnel test section were characterised, $V_{physical}$, for both single-phase and two-phase flows. High resolution in the vertical velocity, compared to \cite{Mora2021settling}, was obtained thanks to a new PDPA setup. This, together with the detailed measurements of alignment and secondary motions, created a more accurate dataset of settling velocity for small Stokes number particles.

The results in this study confirm and extend the trends observed previously (among others by \cite{wang_maxey_1993, Aliseda2002, Good2014, Mora2021settling}). Specifically, the settling velocity enhancement, that has been observed under a wide range of conditions, disappears with an increase of global (wind tunnel) Reynolds number, and turns to hindering at high Reynolds numbers $Re_\lambda > 260$ .
This dependence with Reynolds number is in contradiction with most numerical studies (\cite{Bec2014, Rosa2016, Tom2019}).
However, for a smaller range of Reynolds numbers, the maximum of enhancement is proportional to the inlet velocity $U_\infty$, and therefore to the global Reynolds number. A new phenomenological scaling considering the influence of the bulk velocity has been proposed.

The range of volume fractions investigated is limited, and precludes the influence of this variable on settling enhancement to appear.
Different turbulence generation schemes allow for flows with different integral and Taylor length scales, at the same turbulent intensities and Reynolds numbers. We show that even if the Reynolds number and the turbulent intensity are similar, significant differences in the settling modification remain, due to widely different integral length scales. This suggests an important role of the large flow structures on the settling velocity modification.

\begin{acknowledgments}
This work has been supported by a LabEx Tec21 grant (Investissements d'Avenir - Grant Agreement $\#$ ANR-11-LABX-0030). We also would like to thank Laure Vignal for her help with the PDPA measurements and Vincent Govart for producing experimental rigs. The authors report no conflict of interest. \\

\end{acknowledgments}

\appendix

\section{Additional scalings. \label{A1}}
\begin{figure}[htp]
	\centering 
	\includegraphics[width=\textwidth]{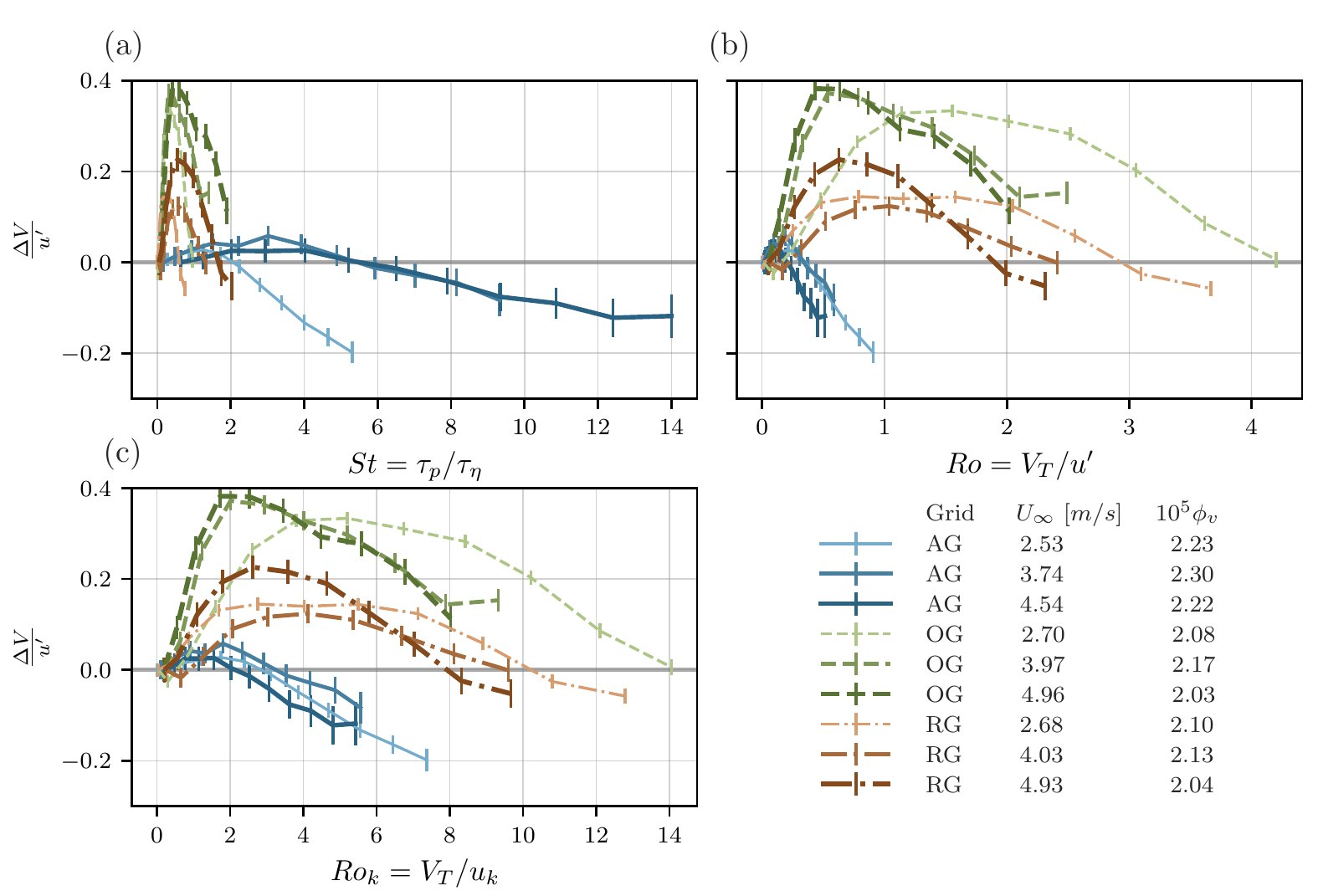}
	\caption{Particle velocity over the carrier phase fluctuations $\Delta V/u'$ against the Stokes number (a), Rouse number (b) and the Rouse number based on the Kolmogorov scale (c) for a volume fraction of $2.0\times10^{-5}$.
	Line styles follow the caption of Figure \ref{fig:DV_vs_D}.
    }
	\label{fig:appendix1}
\end{figure}

Particle settling velocity is often presented against the Stokes number (\cite{wang_maxey_1993, Yang1998, Aliseda2002, Good2014, Rosa2016, Petersen2019, Yang2021}), the Rouse number (\cite{Good2012, Good2014, Mora2021settling}) and a Rouse number based on the Kolmogorov scale $V_T/u_\eta$ (\cite{Good2014}). Figure \ref{fig:appendix1} show the present data against these three different parameters.

\section{PDF of particles' velocities. \label{A2}}
\begin{figure}[htp]
	\centering 
	\includegraphics[width=\textwidth]{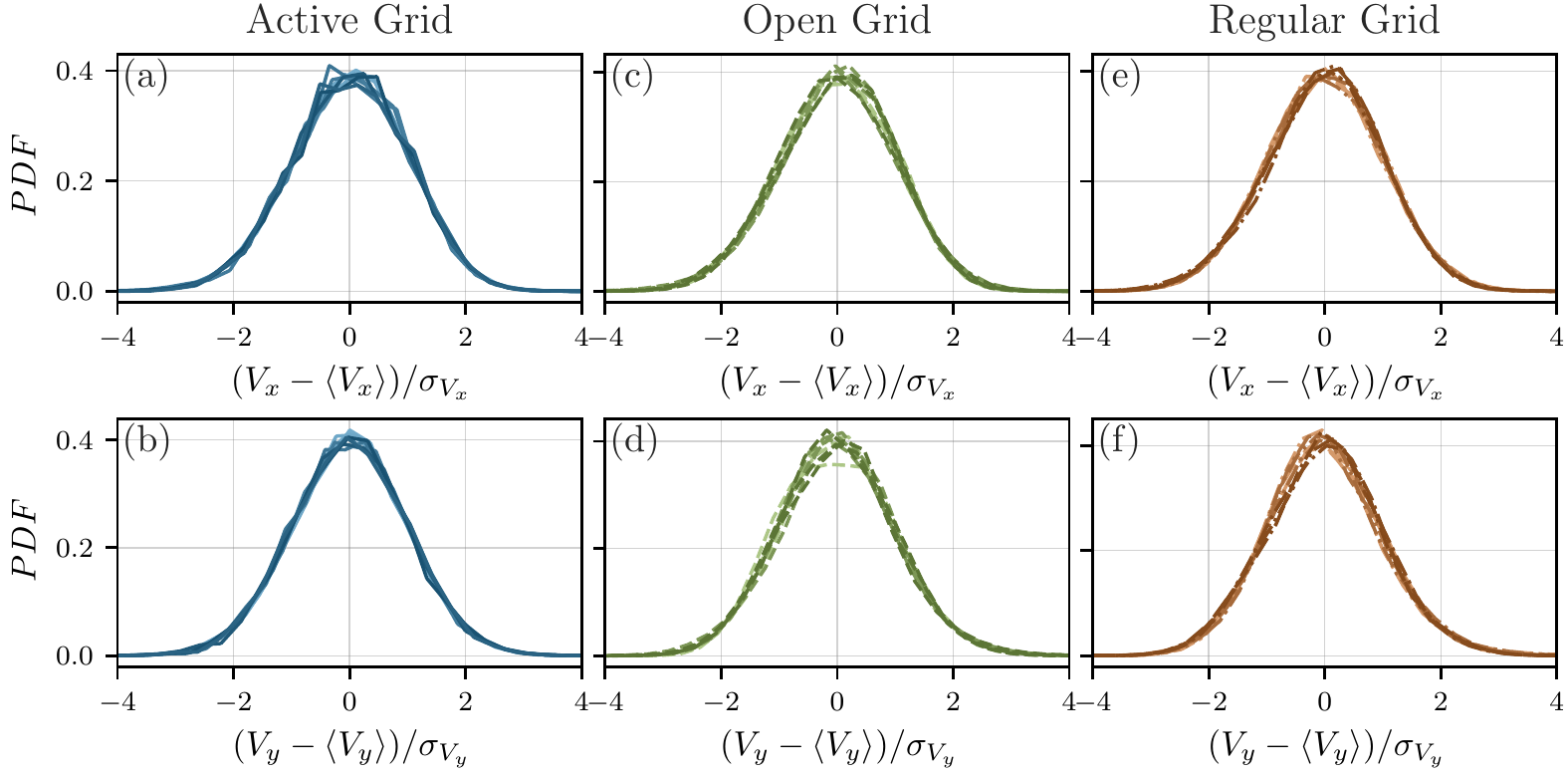}
	\caption{Probability distribution function of the streamwise (top) and vertical (bottom) velocity for each type of grid: active grid (a,b), open grid (c,d) and regular grid (e,f).
    }
	\label{fig:pdf_v}
\end{figure}

In this section we show the raw velocity obtained with the PDPA. It can be observed that all inertial particles horizontal and vertical velocities have a Gaussian distribution (see Figure \ref{fig:pdf_v}).

\section{Determination of the PDPA misalignement angle.}
\label{olive_oil}

A small deviation angle between the PDPA axes and the wind tunnel axes is always present even if the best precautions were taken during the setup of the device.
The deviation angle has a negligible impact on the horizontal velocity but can induce a significant bias on the measurements of the settling velocity, since the particle's horizontal velocity component is much larger than the vertical one.

We call $\beta$ the angle between the axes of the PDPA and the axes of the wind tunnel. $V_{XPDPA}$ and $V_{YPDPA}$ are respectively the streamwise and vertical components of the velocity measured by the instrument while $V_{XWT}$ and $V_{YWT}$ are the exact particle velocity component in the wind tunnel coordinate system (see \cite{Mora2021settling}). 

By projecting the accurate droplet velocity in the frame of reference of the PDPA we get: 

\begin{equation}
	\overrightarrow{V_{YWT}} = \bigg( \underbrace{V_{YPDPA}\cos(\beta)}_{\approx V_{YPDPA}}  - V_{XPDPA}\sin(\beta) \bigg)\overrightarrow{y}
\end{equation}

Since the PDPA was set in noncoincident mode, we do not have access to the horizontal component $V_{XPDPA}$ corresponding to the biased settling velocity.
We then approximate by using the mean of the time series horizontal velocity $V_{XPDPA} \approx \langle U \rangle$ and define the angle-corrected velocity as follow:
\begin{equation}
	V_{YWT} = V_{YPDPA} -  \underbrace{
		\langle U \rangle \sin(\beta)}_{V_{\beta}} \label{eq:corrected}
\end{equation}

In order to compute the vertical velocity due to the horizontal component projection $V_\beta$, we estimated the misalignement angle $\beta$ through measurements of olive oil droplets settling velocities.
We used olive oil to be closer to the limit of very small diameter and very small volume fraction $\phi_v$.
Indeed, olive oil droplets have a much smaller average diameter, $\langle d_p \rangle \approx 3 \mu m$, and a less polydispersed size distribution than water droplets.

The settling velocity of olive oil droplets were collected for different freestream velocities in absence of grid in order to have a flow as laminar as possible. 
Measurements were taken when the probe volume was situated on the center, close to the wall of the wind tunnel and each time the PDPA had to be realigned.
The particle speed in a still fluid is computed from the particle relaxation time $\tau_p$ including the non-linear drag from Schiller and Nauman semi-empirical equation (\cite{Bubblesdrops}):
\begin{equation}
	V_T = \tau_p g \quad \text{with} \quad \tau_p = \frac{\rho_p d_p^2}{18\mu_f(1+0.15Re_p^{0.687})}
\end{equation}
With $\mu_f$ is the air dynamic viscosity, $g$ the gravitational acceleration, $d_p$ the particles' diameters, the oil droplet density $\rho_p = 900 ~ kg.m^{-3}$ and $Re_p =  V_T  D_p/\nu$ the particle Reynolds number.
As the diameter of olive oil droplets is extremely small the actual velocity is supposed to be equal to the Stokes velocity $V_{YWT} = V_T$. We then get from equation \ref{eq:corrected}:
\begin{equation}
	\langle V_{YPDPA} \rangle = V_T + \langle U \rangle \sin (\beta) \label{eq:fit}
\end{equation}

\begin{figure}[htbp]
	\centering \includegraphics[width=0.6\textwidth]{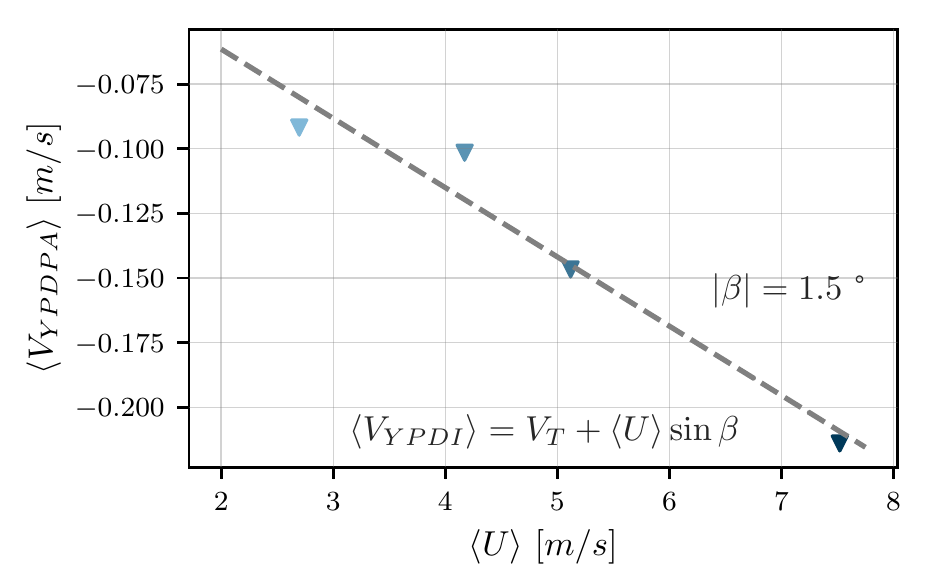} 
	\caption{$\langle V_{YPDI} \rangle $ against $\langle U \rangle$ for the different incoming velocities with olive oil droplets measurements. A linear fit of the data is shown in dashed line. 
	}
	\label{fig:fit}
\end{figure}

With several freestream velocities and equation \ref{eq:fit} a least squares polynomial fit on the values of $\langle V_{YPDPA} \rangle$ and $\langle U \rangle$ can be performed to estimate $\sin(\beta)$. 
Figure \ref{fig:fit} shows $\langle V_{YPDPA} \rangle$  against $\langle U \rangle$ for the probe volume on the center where a linear fit was done and the slope gives the value of $\sin(\beta)$. In our case, $\beta$ is found equal to $\beta = 1.5 \degree \pm 0.3 \degree $.\\

\bibliographystyle{jfm}

\end{document}